\documentclass[preprint,11pt]{elsarticle}

\usepackage{amsmath}
\usepackage[dvips]{epsfig}
\usepackage{graphicx}
\usepackage{amssymb}

\usepackage{setspace}

\usepackage{booktabs}
\usepackage{hyperref}
\usepackage{url}
\usepackage{xcolor}
\usepackage{color}
\usepackage[top=2.5cm, bottom=2.5cm, left=3cm, right=2.5cm]{geometry}

\usepackage{setspace}
\DeclareMathOperator\erfc{erfc}

\journal{European Journal of Mechanics - B/Fluids}

\numberwithin{equation}{section}

\begin{document}

\begin{frontmatter}

\title{Rotation Rate of Particle Pairs in Homogeneous Isotropic Turbulence}


\author[mar,bay]{Abdallah Daddi-Moussa-Ider}
\ead{abdallah.ider@stmail.uni-bayreuth.de}

\author[bay]{Ali Ghaemi}



\cortext[cor1]{Corresponding author}

\address[mar]
 {
 Institut de Recherche sur les Ph\'{e}nom\`{e}nes Hors-\'{E}quilibre\\
  49, Rue F. Joliot-Curie 146, 13384 Marseille, France
  }

\address[bay]
{
 Physikalisches Institut an der Universit\"{a}t Bayreuth \\
 Universit\"{a}tsstra\ss e 30, 95447 Bayreuth, Germany
 }

\begin{abstract}
Understanding  the dynamics of particles in turbulent flow is important in many environmental and industrial applications.
In this paper, the statistics of particle pair orientation is numerically studied in homogeneous isotropic turbulent flow, with Taylor microscale Reynolds number of 300. It is shown that the Kolmogorov scaling fails to predict the observed probability density functions (PDFs) of the pair rotation rate and the higher order moments  accurately. Therefore, a multifractal formalism is derived in order to include the intermittent behavior that is neglected in the Kolmogorov picture. 
The PDFs of finding the pairs at a given angular velocity for small relative separations reveals extreme events with stretched tails and high kurtosis values. 
Additionally, The PDFs are found to be less intermittent and follow a complementary error function distribution for larger separations.

\end{abstract}


\end{frontmatter}

\section{Introduction}

Understanding how particles are advected by fluids is of major interest in many applications, including the environmental and the geophysical flows \cite{shaw03}.
One outstanding example is the eruption of volcanoes, where particles with different sizes and inertia are released into the atmosphere \cite{oberhuber98} and then transported with turbulent currents \cite{mehaddi13, daddi13}. 
In addition to  the environmental importance of understanding the influence of turbulence on particle dispersion, this problem is also of practical interest in the industrial flows, where the advection of particles is involved \cite{lundell11, klinkenberg13}. 

The theoretical studies of the relative separation between two particles are based on stochastic models. 
Many experiments \cite{bourgoin06, ouellette06} and numerical simulations \cite{ishihara02, biferale05b} have been performed over the last few decades in order to accurately evaluate these theories.  
Richardson, in his pioneering work \cite{richardson26}, has examined the relative motion of particle pairs in turbulent flows. 
He has estimated the scale dependency of the eddy-diffusivity coefficient through the observation of dispersed plumes. 
This scale dependency is claimed to be the origin of the accelerated nature of turbulent dispersion \citep{boffetta02}. 
In the inertial range of motion (for $\eta \ll r \ll L$ and $\tau_\eta \ll t \ll T_L$, where $\eta$ is the Kolmogorov dissipative scale, $L$ is the energy injection length scale of the flow, $\tau_\eta$ is the local-eddy-turn-over-time and  $T_L$ is the large time scale), Richardson has suggested that the process of relative dispersion is governed by a diffusion-like equation.
Solving this equation gives the probability of finding a pair of particles at a specific separation,  at any time \cite{salazar09}. 
Nevertheless,  recent studies \cite{biferalearx, scatamacchia12} on the dynamics of tracer  pairs that are released from many point sources have reported severe deviations from the Richardson theory. 
Additionally, this theory does not account for the rotational behavior of particle pairs. 

The orientation dynamics of a single rigid ellipsoid particle under Stokes flow has been described by Jeffery \cite{jeffery22}, where inertia and the thermal fluctuations are neglected. 
The same equation can be used to describe the motion of any axisymmetric particle, provided that its aspect ratio is known \cite{bretherton62}. 
In the presence of weak inertia, Einarsson \textit{et al.} \cite{einarsson15a, einarsson15b} studied the rotation of small and neutrally buoyant axisymmetric particles in a viscous shear flow, by perturbatively solving the coupled particle-flow equations.
Shin and Koch  \cite{shin05} presented the results of direct numerical simulations (DNS) of the translational and rotational motions of  fibers in a fully developed isotropic turbulent flow, for a range of Reynolds numbers.
They concluded that the fibers whose lengths are much smaller than the Kolmogorov length scale $\eta$ translate like fluid particles and rotate like material lines. 
With increasing fiber length, the translational and rotational motions of the fibers slow down as they become insensitive to the smaller-scale eddies.

Using computer simulations, Pumir and Wilkinson  \cite{pumir11} studied the temporal evolution of the orientation vector of microscopic rod-like particles.
They showed that rod-like particles are more strongly aligned with vorticity than with the principal strain axis.
An interesting component of turbulence, which is named the pirouette effect, has been reported by Xu \textit{et al.} \cite{xu11}.
It has been shown that the axis of rotation of tetrahedra tracers aligns with the initially strongest stretching direction, a phenomena which can be justified by the conservation of the angular momentum.
Using video particle tracking technique, Parsa \textit{et al.} \cite{parsa11} made accurate simultaneous measurements of the motion and the orientation of the rods in the presence of a bi-dimensional chaotic velocity field.
The first three-dimensional experimental measurements of the orientation dynamics of rod-like particles was also reported by Parsa \textit{et al.} \cite{parsa12}. 
In this work, a good agreement with the previous numerical simulations was reported. 
Moreover, Parsa and colleagues   \cite{parsa14} studied the rotation rate of the rods with lengths in the inertial range of turbulence. 
They presented the experimental measurements of the rotational statistics of neutrally buoyant rods, and derived a scaling law for the mean-squared rotation rate. The latter showed a good agreement with the Kolmogorov classical scaling.

It is noteworthy that, to the best of our knowledge, the orientation of the tracer pairs has not been studied so far.
Therefore, this question will be addressed here, by computing the probability of finding a  pair of particles at a specific rotation rate, given the relative separation between them. 
By considering a particle pair with a specific separation, one can study the orientation statistics of an imaginary tracer rod that is delimited by these two particles.
Furthermore, for the first time, the pair rotation rate PDFs at  later times will be presented in this work.

This paper is organized as it follows. 
The next section describes the numerical data, and the approach that is used to derive the rotation rate of the tracer pairs. 
Afterward, the intermittency behavior that is observed at the small scales of motion is discussed in detail.
In Sec. \ref{resultsanddiscussions}, the multifractal (MF) formalism to intercept the probability density function (PDF) distributions is presented.
Thereafter, the   higher-order moments of the rotation rate are evaluated  and compared to the multifractal prediction together with the Kolmogorov (K41) picture.
Finally, the concluding remarks, and the outlooks are stated in Sec. \ref{sectconc}.

\section{Methods}\label{methods}

\subsection{Particle pair orientation}
The rotational dynamics of particle pairs in turbulent flow can be addressed by solving the dimensionless incompressible Navier-Stokes equations for a Newtonian fluid, i.e.
\begin{eqnarray}  
 \frac{\partial \mathbf{u}}{\partial t} +  \mathbf{u} . \boldsymbol{\nabla} \mathbf{u}
  &=& -\boldsymbol{\nabla} p + \frac{1}{\operatorname{Re}} \boldsymbol{\nabla}^2 \mathbf{u} + \mathbf{f}, \label{NS1}
  \\
  \nabla . \mathbf{u} &=& 0,\label{NS2}
\end{eqnarray}
using Direct Numerical Simulations (DNS).
In these equations, $\mathbf{u}$ denotes the tri-dimensional velocity field, $p$ is the pressure and Re is the flow Reynolds number, $\operatorname{Re} \equiv L u'/\nu$.
Re measures the ratio between the nonlinear inertial forces, and the linear viscous forces, 
where $u'$  is the root-mean-squared velocity and $\nu$ is the fluid kinematic viscosity which is defined as the ratio between the dynamic molecular viscosity $\mu$ and, the fluid density $\rho$. 
In order to avoid energy dissipation, the flow was forced by  keeping the total energy constant in the first wavenumber shells,  by applying a large-scale forcing term $\mathbf{f}$.
This force injects energy at a mean rate of $\epsilon = \langle  \mathbf{f}.\mathbf{u} \rangle$, where $\epsilon$ is the mean turbulent kinetic energy dissipation rate \cite{daddi14report}.
The integration of the equations of motion  is performed on $1,024^{3}$ regular cubic lattice with Taylor microscale Reynolds number of 300, and periodic boundary conditions.
The simulation parameters are shown in Tab. \ref{simulationpara}.
With the present choice of parameters, the dissipative range of length scales is well resolved because the grid size $\Delta x$ is in the range of Kolmogorov length scale $\eta$ (as it is reported in more detail by Biferale \textit{et al.} \cite{biferalearx}).
A  fully-de-aliased parallel pseudospectral code, with a second-order Adams-Bashforth temporal scheme, for 3D homogeneous isotropic turbulence, assuming constant fluid viscosity and density, is used in order to solve Eqs. \eqref{NS1} and \eqref{NS2}.

The fluid is seeded with bunches of tracers emitted within a small region which has a size comparable to the Kolmogorov dissipative scale $\eta $.
The emission is carried out in puffs of 2,000 particles that are followed during a maximum time of 160 $\tau_\eta$.
The tracer particles take on the fluid velocity immediately, and adapt the rapid fluid velocity fluctuations \cite{klein12}.
Therefore, the velocity of each tracer is related to the instantaneous fluid velocity by the following equation
\begin{equation}
 \mathbf{v_p} \equiv \frac{d \mathbf{x_p}}{dt} = \mathbf{u} (\mathbf{x_p}(t), t). 
 \label{lagrangian}
\end{equation}

The particle trajectories are computed by integrating the Eq. \eqref{lagrangian}.
The positions and the velocities of each particle are stored at a sampling rate of $\tau_\eta$.
For the 2,000 particles generated in each puff, all the possible pairs (approximately two millions) are considered \cite{daddi14}.

\begin{table}
\begin{center}
\caption{Parameters of the numerical simulation.
\label{simulationpara}}

\vspace{.15cm}

\begin{tabular}{lr}
\toprule[1.5pt]
Kolmogorov dissipative active length scale $\eta$&0.005\\
\hline
Spacing between two collocation points in the regular cubic lattice $\Delta x$&0.006\\
\hline
Mean turbulent kinetic energy dissipation rate per unit mass $\epsilon$&0.81\\
\hline
Fluid kinematic viscosity $\nu$&0.00088\\
\hline
Dissipative time scale $\tau_\eta$&0.033\\
\hline
Integral time scale $T_L$&67\\
\hline
Energy injection length $L$&2$\pi$\\
\hline
Number of collocation points $N$&1,024\\
\hline
Turbulent velocity fluctuation $u'$&1.7\\
\bottomrule[1.25pt]
\end{tabular}

\end{center}
\end{table}

\subsection{Pair angular velocity}

The pair angular velocity can be measured on the basis of the instantaneous positions and the velocities of the two particles.
Consider a given pair of particles defined at every time by its separation vector $\mathbf{r}$ pointing from the first tracer particle toward the second one. 
One of the particles  is taken as a reference point for computing the pair orientation. 
The relative velocity of the other particle $\Delta \mathbf{u}$ is decomposed into two components;
i.e. $\mathbf{\Delta u_\parallel}$ parallel to the separation vector, and $ \Delta \mathbf{ u_\perp}$ perpendicular to it. 
The first particle and the transverse component of the relative velocity of the second one, defines a plane of rotation.
The axis of rotation ${\mathbf{e}}$ is then normal to this plane, and defines the direction of the angular velocity pseudovector $\boldsymbol{\omega}$.
By taking $\theta$ as the angle between the separation vector $\mathbf{r}$ and $\Delta \mathbf{ u}$, then the angular velocity vector can be written as
\begin{equation}
 \boldsymbol{\omega} 
 =\frac{|\Delta  \mathbf{u_\perp}| }{|\mathbf{r}|}{\mathbf{~e}}
 = \frac{|\Delta  \mathbf{u}| \sin \theta}{|\mathbf{r}|}{\mathbf{~e}}, 
\end{equation}
or
\begin{equation}
 \boldsymbol{\omega} = \frac{  \mathbf{r} \times \Delta  \mathbf{ u} }{r^2}.
\end{equation}
and in tensorial notation,
\begin{equation}
 \omega_k=\frac{\epsilon_{ijk} r_i \Delta u_j}{r^2},
 \label{angveldef}
\end{equation}
where $\epsilon_{ijk}$ is the Levi-Civita  symbol. 
Considering the isotropy of turbulence, the ensemble average over the angular velocity components vanishes, that is to write $\langle \omega_k \rangle = 0$.

\section{Results and discussions}\label{resultsanddiscussions}

\subsection{Multifractal model}

The angular velocity of the rotating particle pairs in Eq. \eqref{angveldef} can be computed at any time by knowing the  spatial positions, and the velocities of the two tracer particles separated by a vector $\mathbf{r}$.
In the case of inertialess particles, the angular velocity can be viewed as the ratio between the  Eulerian transverse velocity increment $ \delta_r u_\perp \equiv u_\perp(r + \delta r) - u_\perp(r)$, and the instantaneous distance $r$.
The mean-squared angular velocity $\langle \omega^2 \rangle $ for a fixed separation $r$, is thus the ratio between the second order Eulerian transverse velocity structure function, and $r^2$.

For the purpose of computing the higher-order moment of angular velocity that is parametrized by  $r$, 
a series of computations over the whole number of bunches at different times, and for different constant separations are performed.
In the inertial range of motion, the size of the eddies that move the pairs apart varies with the separation distance.
The eddies with the scales in the order of $r$ are the most effective in the dispersion process \citep{corrsin62}.
Intuitively, in this intermediate range,
the further the separation between the two particles is, the slower the pair will rotate.
This happens because in the Kolmogorov phenomenology, the angular velocity modulus scales as $r^{-2/3}$. 
According to the Kolmogorov second similarity hypothesis \cite{kolmogorov41}, all statistically averaged quantities at scale $r$ in the intermediate range are uniquely determined by $\epsilon$ and $r$. 
The  $p$th order moment  angular velocity should then scale in the inertial range as 
\begin{equation}
 \langle \omega^p \rangle \simeq {r^{-2p/3}}.
 \label{omegar}
\end{equation}

However, in the dissipative range of turbulence (i.e. for $r \ll \eta$), the higher order moments are independent of the scale $r$.
A suitable expression is therefore needed in order to fulfill the requirements of both ranges.
In Kolmogorov phenomenology, the higher order moments are expressed with respect to separation $r$ via the Batchelor parametrization \cite{batchelor51}, which satisfies both the dissipative range and the inertial range.
Then $\langle \omega^p \rangle$ is expected to be scaled as \cite{chevillard06}
\begin{equation}
 \langle \omega^p \rangle  \simeq \frac{\left( \frac{r}{L} \right)^{-2p/3}}{\left( 1+\left( \frac{r}{\eta} \right)^{-2} \right)^{p/3} },
 \label{batchelorparaK41}
\end{equation}

For separations in the inertial dissipative range (i.e. $r > \eta$ and $r \lll L$), these moments undergo a quadratic decay.

Kolmogorov theory  fails to accurately predict the intermittency behavior, because it neglects the presence of fluctuations in the energy transfer from the large scales toward the small ones \cite{parisi85}.
In order to include the intermittency effect in the energy cascade, a multifractal formalism is used together with a  generalization of the Batchelor parametrization,  to describe the  higher order  moments of the angular velocity for both the inertial dissipative range and the inertial range \cite{meneveau96}.
The multifractal approach to turbulence is based on the assumption that the statistical properties of turbulent flows do exhibit scaling properties even if there is intermittency \citep{benzi09b}.
Under these assumptions, the transverse velocity increment reads \cite{chevillard06}
\begin{equation}
\delta_r u_\perp  = v_0 \cfrac{\left( \frac{r}{L} \right)^{h}}{\left(  1+\left( \frac{r}{\eta(h)} \right)^{-2} \right)^{(1-h)/2}},
\label{trans}
\end{equation}
where $v_0$ denotes the large scale transverse velocity fluctuation and $h$ is the singularity in the MF phenomenology.
In this picture, the viscous scale is not a fixed homogeneous quantity, but it fluctuates wildly due to intermittency \citep{benzi10}.
The Kolmogorov length scale $\eta $ is a function of the singularity $h$, and it is related to the Reynolds number by $\eta / L \propto \operatorname{Re}^{-1/(1+h)}$.

Following  Eqs. \eqref{angveldef} and \eqref{trans}, the angular velocity component at a distance $r$ is estimated in the MF model as
\begin{equation}
\omega \equiv \frac{\delta_r u_\perp}{r} \sim 
\frac{v_0}{L}
\cfrac{\left( \frac{r}{L} \right)^{h-1}}{\left(  1+\left( \frac{r}{\eta(h)} \right)^{-2} \right)^{(1-h)/2}},
\label{omega}
\end{equation}

A generalization of the probability to observe an Eulerian velocity fluctuation at scale $r$ with a local H\"{o}lder exponent $h$,  is given by \cite{chevillard06}
\begin{equation}
 P_h(r)\propto \cfrac{\left( \frac{r}{L} \right)^{3-D(h)}}{\left( 1 + \left( \frac{r}{\eta(h)} \right)^{-2} \right)^{(D(h)-3)/2}},
 \label{P(r)}
\end{equation}
where $D(h)$ is the spectrum of the fractal dimension.

\subsection{Probability density function distribution}\label{probabilityDensityFunctionDistribution}

In this study, there are many particle pairs which are embedded in the turbulent flow.
The most comprehensive and accurate description of the orientation state can be achieved by evaluating  the PDFs of the orientation of the pairs.
The non-Gaussianity is one of the fundamental features of 3D turbulence with a non-zero skewness, and a very large kurtosis of the PDF distributions \citep{benzi91}.
A superposition of stretched exponentials corresponding to the various singularity exponents are usually observed at small scales.
In the following, we shall be interested in the PDF  of finding a pair of tracers at a given angular velocity component $\boldmath{\omega}$ when it reaches a given separation $r$.

The MF phenomenology is capable of predicting the PDFs of the velocity increment and  the accelerations in both the Eulerian and Lagrangian frames \cite{toschi05, toschi09}.
The MF model can also successfully intercept the numerically  observed angular velocity PDFs.
In this section, an MF model is employed for the PDFs of the angular velocity components.
The PDFs are allowed to take values in all the axis of reals.
Since the flow is isotropic,  i.e. the pair properties are distributed independently of orientation,
the distributions are symmetric.
Furthermore, the PDFs that correspond to each of the components all collapse on a single curve.

By assuming that the large scale velocity field $\mathbf{v_0}$ is Gaussian with $d$ components, its magnitude has the following PDF distribution \cite{biferale08}
\begin{equation}
 P(v_0) d v_0  = v_0^{d-1} \exp \left( -\frac{v_0^2}{2} \right),
\end{equation}
where  the normalization constant has been omitted here.
The transverse velocity modulus corresponds to $d = 2$, and the PDF follows a Rayleigh distribution.
By setting the variance to unity, one can write the large scale transverse velocity  PDF modulus as
\begin{equation}
 P(v_0) = v_0 \exp\left(-\frac{v_0^2}{2}\right),
 \label{Pm(v)}
\end{equation}

\begin{figure}
\centering
    \scalebox{1.08}{\input{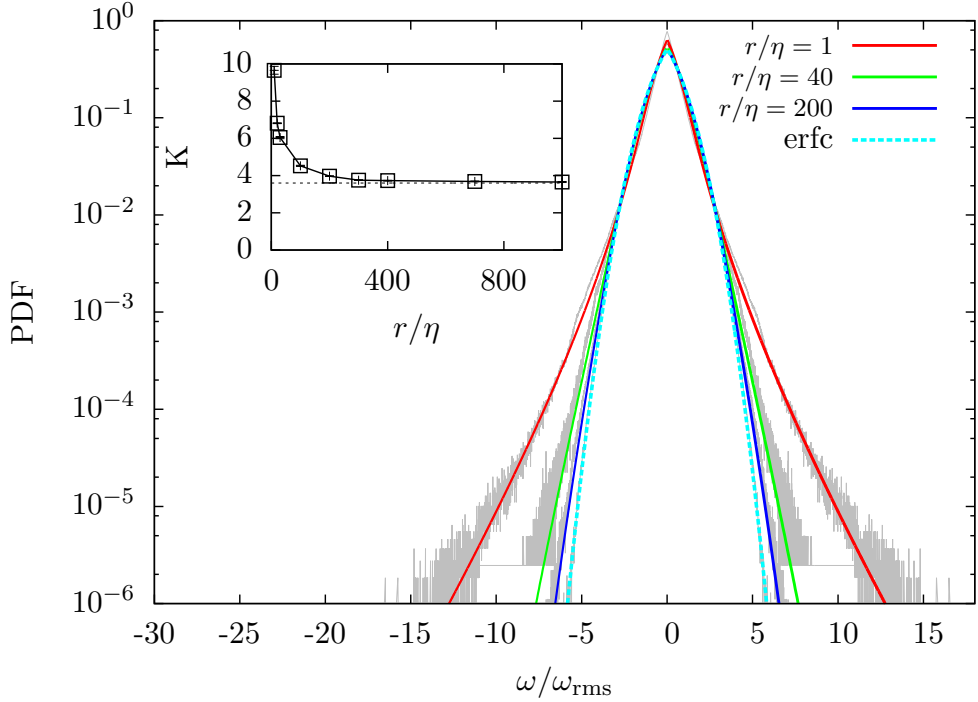}}
    \caption{(Color online) A comparison between the numerical results (in gray) and the MF prediction of  the normalized PDFs of the angular velocity component, that are calculated for the separation lags of $r/\eta$ = 1, 40, 200 and 1,000. 
    The dashed line represents the unit variance complementary error function given by \eqref{P(v)}.
    It can seen that the complementary error function overlaps perfectly on the numerical results for $r = 1,000 \eta$.
    Inset:
    The variations of the kurtosis of the angular velocity component PDF.
    The horizontal dashed line corresponds to 18/5 as the kurtosis of the complementary error function.}
    \label{kompo}
\end{figure}

It is possible to derive the PDF of the large scale transverse velocity component fluctuation from the PDF of the transverse velocity modulus given by \eqref{Pm(v)}.
Because the three components have the same probability distribution due of isotropy, the PDF of each of them is  given by a complementary error function (see the appendix for more  details). For a variance of unity the following large scale transverse velocity component PDF is derived
\begin{equation}
 P(v_0) = \frac{\sqrt{\pi}}{2\sqrt{3}}   \erfc\left(\frac{|v_0|}{\sqrt{3}}\right).
 \label{P(v)}
\end{equation}

The probability of finding a pair of particles at a specific angular velocity, and for a given separation $r$ is obtained from
\begin{equation}
 P(\omega | r) = 
 \int\limits_{h \in I}
 { P(\omega | h) P_h(r)}~\mathrm{d}h.
 \label{PP}
\end{equation}

By plugging Eqs. \eqref{P(v)} and \eqref{P(r)} into \eqref{PP}, and using the probability conservation law for the equation \eqref{omega}, one can write
\begin{equation}
P(\omega | r) \propto 
\int\limits_{h \in I}
\left(  \left( \frac{r}{L} \right)^{2}   + \left( \frac{\eta(h)}{L} \right)^{2} \right)^{2-\frac{D(h)+h}{2}}
\erfc 
\left(   
\frac{L |\omega|}{\sqrt{3}} \left(  \left( \frac{r}{L} \right)^{2}   + \left( \frac{\eta(h)}{L} \right)^{2} \right)^{\frac{1-h}{2}}
\right).
\label{probacompo}
\end{equation}

The fractal dimension has the following form in a log-Poisson process  \citep{she94, dubruelle94}
\begin{equation}
 D(h) =  \frac{3(h-h_0)}{\ln \beta}   \left[    \ln \left(  \frac{3(h_0 - h)}{d_0 \ln \beta}  \right) -1  \right] +3 -d_0,
\end{equation}
where $\beta = 0.6$, $h_0 = 1/9$ and $d_0 = (1-3h_0)/(1-\beta) = 5/3$.
The lowest singularity $h_{min} = 1/9$ is obtained by considering $\lim_{h \to h_{min}} D(h) = 0$. 
The fractal dimension attains its maximum which is $D=3$, for $h_{max} = h_{peak} \simeq 0.38$
\citep{benzi10}.

In Kolmogorov theory, the singularity $h$ takes only the value of 1/3, and the fractal dimension is restricted to $D=3$.
It is possible to derive the PDF distribution in this particular case from Eq. \eqref{probacompo}.
After computation, it is found that the angular velocity component PDF is given by
\begin{equation}
 P(\omega | r) = \frac{1}{\omega_{\text{rms}}} \frac{\sqrt{\pi}}{2 \sqrt{3}}  \erfc \bigg( \frac{1}{\sqrt{3}} \frac{|\omega|}{\omega_{\text{rms}}} \bigg),
\end{equation}
where $\omega_{\text{rms}}$ is the root-mean-squared angular velocity, expressed in Eq. \eqref{batchelorparaK41} for $p=2$.
The complementary error function distribution given by \eqref{P(v)}, is therefore recovered for the normalized PDF.

\subsubsection{Comparison with numerical results}

In Fig. \ref{kompo}, a comparison between  our numerical results and the MF prediction given by \eqref{probacompo} is presented.
The agreement between these two  shows the robustness of the MF approach in reproducing the DNS results accurately.
The PDF of the angular velocity component for the smallest separations i.e. of the order of the Kolmogorov length scale $\eta$, is highly  intermittent and the PDF has stretched tails that extend beyond  thirty times the standard deviation of angular velocity (for the sake of convenience, in Fig. \ref{kompo} tails are not shown in full lengths).
This indicates  that when the separation length $r$ is of the order of the smallest scale of the flow, the probability of high rotation rate events are much higher than those predicted by a complementary error function distribution.
An increasingly large number of particles is needed in order to ensure a sufficient statistical convergence of the data, and therefore to  describe  the long tails accurately.
The inset of Fig. \ref{kompo} illustrates this by showing the variations of the PDF kurtosis as a function of $r$. 
The kurtosis is a measure of the distribution pickedness.
And for symmetrical distributions, it is defined as the ratio between  the fourth order moment, and the square of the distribution variance.
That is to write,
\begin{equation}
K \equiv \frac{\langle \omega^4 \rangle}{\langle \omega^2 \rangle^2}.
\label{kur}
\end{equation}

For small separations, the kurtosis is very large and it reaches a value of 10 for a separation of the order of the Kolmogorov scale. 
The larger the separation is, the smaller the kurtosis will be, and it finally converges toward 18/5.
This is the kurtosis of the large scale angular velocity component PDF given by \eqref{P(v)} 
(see the appendix for more details about the computation of this kurtosis).

\subsection{Higher order moments}

\begin{figure}
\centering
    \scalebox{0.7}{\input{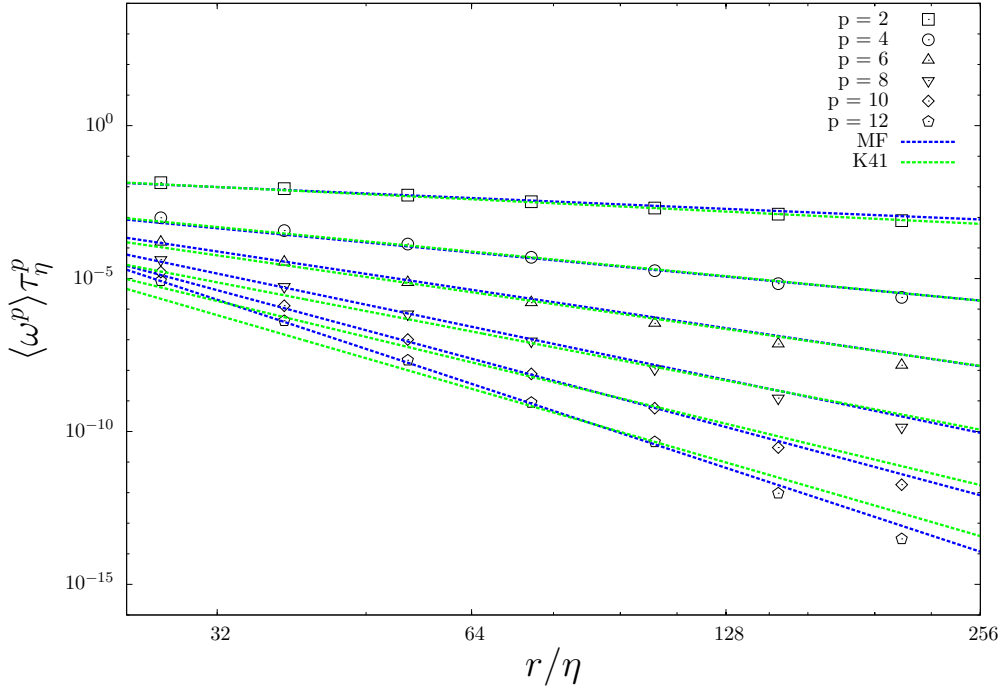}}
    \caption{(Color online)
    Comparison  between the Kolmogorov scaling and the multifractal model.
    The scaled higher order moments of the angular velocity component is depicted  versus the scaled separations in the inertial range.
 For $p = 2, 4$, the MF model and the K41 scaling given by $r^{-2p/3}$ are almost indistinguishable.
 The  multifractal prediction shows a better agreement with the numerical results for the higher order moments $p=6, 8, 10, 12$.
 }
    \label{moment2}
\end{figure}

The $p$th  order moment of the angular velocity can be computed from the PDF distributions as
\begin{equation}
    \langle \omega^p \rangle = 
    \int\limits_{\omega \in \mathbb{R}}
    { \omega ^p P(\omega | r)}\mathrm{d} \omega.
    \label{HOM}
\end{equation}

By plugging Eqs. \eqref{probacompo} into \eqref{HOM}, and integrating $\omega$ over the whole axis of reals, the integral expression of the  higher order moment of the angular velocity component at separation $r$ is
\begin{equation}
 \langle \omega^p \rangle \simeq 
 \int\limits_{h \in I}
\left(  \left( \frac{r}{L} \right)^{2}   + \left( \frac{\eta(h)}{L} \right)^{2} \right)^{\frac{3}{2} - \frac{1}{2} \left( D(h) + p(1-h) \right)}
 \mathrm{d}h.
\end{equation}

When the support $I$ is limited to $h = 1/3$ with a fractal dimension of $D = 3$, the Kolmogorov scaling given by \eqref{batchelorparaK41} is recovered.
Eq. \eqref{HOM} can also be written as follows, by integrating with respect to the singularity $h$
\begin{equation}
 \langle \omega^p \rangle \simeq \int\limits_{h \in I} \omega^{p} P(\omega | h) \mathrm{d} h.
\end{equation}

In Fig. \ref{moment2} the scaled higher order moments of the angular velocity component are plotted against the scaled separations $r$ that extend from the inertial range deep into the dissipative inertial range.
For the first two even order moments, the Kolmogorov picture and the MF prediction are almost indistinguishable. 
In fact,  the intermittency behavior can only be observed for the higher order moments, for which the simulation results present a steeper scaling.

Nevertheless, the Kolmogorov prediction given by \eqref{batchelorparaK41} cannot accurately reproduce the numerical results for  the higher moments  beyond the sixth order moment.
Additionally, the  normalized PDFs that are computed using this theory, for different separations all overlap to the same graph.
In other words, K41 can only be applied for the large separations, when the complementary error function distribution is adopted.
In contrast,  the multifractal model is found to be in a satisfactory agreement with the simulation results for all the separation ranges.

Assuming that in the proposed MF model, the $p$th order moment of the angular velocity in the  inertial range, can be written as
\begin{equation}
 \langle \omega^p \rangle \simeq {r^{-\frac{2}{3}p + \tau_p}},
 \label{intermittencyCorrections}
\end{equation}
that is to write the scaling exponent as the sum of the exponent which is predicted by the Kolmogorov theory in Eq. \eqref{omegar}, and a correction $\tau_p$ which takes into account the intermittency effects.
For the higher order moments, the correction has a negative contribution in the exponent, and the MF model leads to a steeper scaling, giving a better agreement with the numerical results.
This happens because MF includes the intermittency character of the energy dissipation which is neglected in the classical theory.

\begin{figure}
\centering
     \scalebox{1.05}{\input{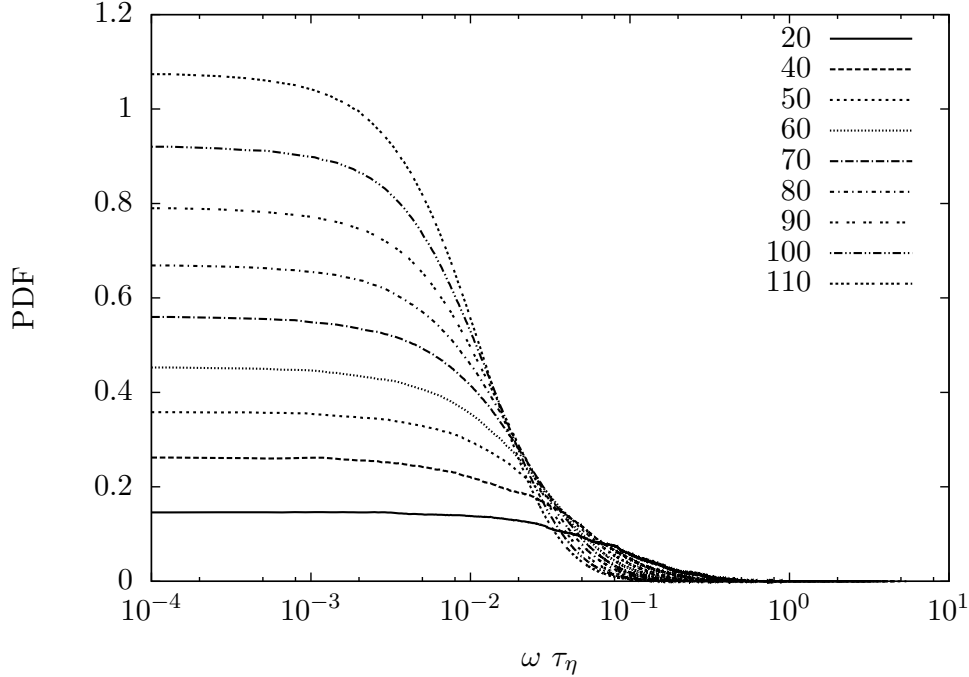}}
     \caption{Lin-log plot of the positive part of the symmetrical PDF of finding the pair of tracers at a given scaled angular velocity, in the range extending from 20 $\tau_\eta$ (solid line) to 110~$\tau_\eta$.
     In this figure, $\omega$ is normalized by the inverse of the eddy-turn-over time  at the Kolmogorov scale $\tau_\eta$.
     At later times, the bunches experience more expansion, and the probability of finding the pairs at a lower rotation rate becomes higher.
     }
     \label{pdftime}
\end{figure}

In Fig. \ref{pdftime} the PDF distribution of the pair angular velocity components at various times is illustrated.
At the initial time, the particles are located at a source with dimensions of the order of the Kolmogorov scale $\eta$.
Scatamacchia \textit{et al.} \cite{scatamacchia12, Biferale14} reported an abrupt transition in the particle dispersion at a time of about $10\tau_\eta$ after the bunch emission. 
At that time, most of the pairs reach a relative distance of the order of $10\eta$.
At times larger than $20 \tau_\eta$, the relative separation of the majority of pairs is within the inertial range of turbulence.
In this range of motion, the angular velocity is scale dependent, and it is inversely proportional to the distance separating the two particles.
This means that at later times, the probability of finding the pair at a smaller angular velocity is much higher, because the bunch of particles has experienced a larger dispersion.
The maximum of the PDF distribution increases with time and the region of small rotation rates becomes more populated.

\section{Conclusion}\label{sectconc}

In this work,
the orientation statistics of two particles as they are carried in a flow are studied.
The flow is seeded by  tracer particles that are released from one point-like source in bunches of thousands.
Knowing the instantaneous positions and the velocities of the emitted particles,
the higher order moments of the angular velocity component  of the rotating pairs are measured.
The theory of Kolmogorov based on purely dimensional arguments is found unable to accurately fit the numerical results.

The multifractal phenomenology is therefore employed in order to include the intermittency corrections.
It is shown that it could remarkably reproduce  the observed behaviors for the angular velocity higher order moments and for the orientation PDFs.
The computation of the PDFs of finding the particle pairs at a specific angular velocity component when they reach a given separation  reveals that one can quite easily attain events in excess of thirty times the standard deviation with high kurtosis values.
The dramatic change of the angular velocity PDF shape provides with immediate impression of how strongly intermittent rotation events are at small scales of motion.
By increasing the separations, the PDFs become less intermittent and converge toward the complementary error function distribution.

Here, the main attention is restricted to the neutrally buoyant particles which are passively transported by the turbulent flow. 
The role of particle inertia on pair orientation should be explored in the future.
It is important to note that the measurements in the  database are not completely  unbiased, because only the particles that are emitted from a single point source are considered.
In order to reduce the spatial correlation, it is worth considering the pairs which are formed from various point-like  sources.
For reliable statistics, a large number of particles will be required. 
It is worth mentioning that an ideal framework for the study of the orientation statistics is a 2D turbulence, because of the absence of intermittency.
Similar measurements and quantitative investigations can be carried out as well in this particular case, where the angular velocity vector has a unique component along the axis normal to the computational domain. 

\section{Acknowledgment}

The first author would like to thank Prof. Federico Toschi, Prof. Luca Biferale,  and Dr. Riccardo Scatamacchia for supervising his thesis and for providing the numerical data,
in addition to their useful discussions and helpful comments on the manuscript during its preparation.
Furthermore, he gratefully acknowledges the support from Eurasmus mobility program. 
He also thanks Dr. Behnam Farid and Dr. Renaldas Urniezius for the online discussions via ResearchGate.


Author contributions: A.D.M.I. performed research and analyzed data; A.D.M.I. and A.G. wrote the paper.
The authors declare no conflict of interest.

\section*{Appendix}
\addcontentsline{toc}{chapter}{Appendices}

Here,  the PDF of the components are derived for a given vectorial quantity, by knowing the PDF of the modulus when the isotropy is guaranteed. 
In the following, $x$, $y$ and $z$ represent the vector components.
Let  $N(r)$ be the  distribution function of the variable $r$ where $r^2 = x^2+y^2+z^2$.

$N(r)$ is normalized to unity, that is to write
\begin{equation}
 \int_{\mathbb{R}^3} N\Big((x^2 + y^2 + z^2)^{1/2}\Big) \mathrm{d}x \mathrm{d}y \mathrm{d}z = 1,
 \label{r3}
\end{equation}

Using the spherical coordinates, Eq. \eqref{r3} can be written as,
\begin{equation}
 \int_{0}^{\infty} 4 \pi r^2 N(r) \mathrm{d}r = 1.
\end{equation}

The  PDF for the modulus $Q(r) = 4 \pi r^2 N(r)$ has a total weight of unity.
Because of isotropy, the PDFs are all the same along all the three axis.
Then only $P(z)$ will be calculated here.

\begin{equation}
 P(z) =  \int_{\mathbb{R}^2} N \Big((x^2 + y^2 + z^2)^{1/2}\Big) \mathrm{d}x \mathrm{d}y,
 \label{P}
\end{equation}

$P(z)$ can be written as a function of $Q(r)$, 
\begin{equation}
 P(z) =  \frac{1}{4\pi} \int_{\mathbb{R}^2} \frac{Q\Big( (x^2 + y^2 + z^2)^{1/2}\Big)}{x^2 + y^2 + z^2} ~\mathrm{d}x \mathrm{d}y,
\end{equation}

If the two-dimensional vector whose Cartesian coordinates are $x$ and $y$ are described in the cylindrical coordinates ($u$, $\phi$, $z$), where $u$ varies over $[0,\infty[$ and $\phi$ over $[0,2\pi]$, then  $u^2 = x^2+y^2$, and

\begin{equation}
 P(z) = \frac{1}{2} \int_{0}^{\infty}  \frac{Q\Big((u^2+z^2)^{1/2}\Big)}{u^2+z^2} ~u \mathrm{d}u,
\end{equation}

By considering the change of variables $t^2 = u^2+z^2$, then $t\mathrm{d}t = u\mathrm{d}u $, 
the following PDF for the positive values of $z$ will be obtained,

\begin{equation}
 P(z) = \frac{1}{2} \int_{z}^{\infty} \frac{Q(t)}{t} \mathrm{d}t.
 \label{result}
\end{equation}

The resulting equation \eqref{result}  can be illustrated in one well-known  example of turbulence.
It is known that the large scale velocity has  the following distribution,
\begin{equation}
 Q(t) = \frac{4}{\sqrt{\pi}} t^2 e^{-t^2}.
\end{equation}

Using Eq. \eqref{result} one finds that the component PDF has a Gaussian distribution,
\begin{equation}
 P(z) = \frac{1}{\sqrt{\pi}} e^{-z^2},
\end{equation}
which has  a total weight of unity.

As it is discussed in \ref{probabilityDensityFunctionDistribution} , the large scale transverse velocity fluctuation has a Rayleigh distribution given by Eq. \eqref{Pm(v)}, then the PDF of each component has a complementary error function distribution  after integrating Eq. \eqref{result},
\begin{equation}
 P(z) = \frac{\sqrt{2\pi}}{4} \erfc\left( \frac{|z|}{\sqrt{2}} \right).
\end{equation}
which has also a total weight of unity.

Moreover, using Eq. \eqref{kur}  the kurtosis of the above distribution is calculated as
\begin{equation}
 K = \frac{1}{\pi} \left. {\int_0^\infty z^4 \erfc \left( \frac{z}{\sqrt{2}} \right) }
 \middle/
 {\left( \int_0^\infty z^2 \erfc\left( \frac{z}{\sqrt{2}} \right) \right)^2} \right. =\frac{18}{5}.
\end{equation}


\bibliographystyle{elsarticle-num} 


\end{document}